\documentclass[11pt]{article}

\usepackage{amsmath}
\usepackage{amssymb}
\usepackage{times}
\usepackage{dsfont}
\usepackage{theorem}
\usepackage{graphicx}

\def\=d{\,{\buildrel\rm def\over =}\,}

\newcommand{\lambdabar}{{\mkern0.75mu\mathchar '26\mkern -9.75mu\lambda}}

\title{Localization properties and causality aspects\\
of massless and massive scalar particles}
\author{Andreas Walter Aste$^{a,b},$ Maik Uwe Frensel$^a$\\
$\quad$\\
$^{a}$\emph{Department of Physics, University of Basel, 4056 Basel, Switzerland}\\
$^{b}$\emph{Paul Scherrer Institute, 5232 Villigen PSI, Switzerland}}

\date{October 16, 2015}

\begin{document}
\maketitle
\setlength\parindent{0pt}

\begin{abstract}
Localization properties of scalar single particle states are analyzed by explicit calculational
examples with a focus on the massless case. Problems arising from the
non-existence of relativistic particle position operators respecting
the causal structure of Minkowski spacetime are illustrated by exploring the
conflicts arising from localization and causal properties commonly imposed on single particle
states. These topics necessitate the introduction of quantum field theoretical localization concepts
and are scarcely discussed and often misinterpreted in the literature.
\\
\vskip 0.1 cm {\bf Physics and Astronomy Classification Scheme (2010).} 11.10.-z - Field theory;
11.30.-j Symmetry and conservation laws.
\end{abstract}

\section{Introduction}
One might think that everything has been said about the Klein-Gordon equation \cite{Klein,Gordon,Pauli_Weisskopf}
\begin{equation}
\biggl( \frac{1}{c^2} \frac{\partial^2}{\partial t^2} - \Delta  +  \lambdabar_c^{-2} \biggr) \varphi(ct,\vec{x})=
(\Box + \lambdabar_c^{-2}) \varphi(x) = 0 \, , \quad \lambdabar_c = \frac{\hbar}{m c}
\label{Klein-Gordon_equation}
\end{equation}
or about the simpler wave equation
\begin{equation}
\Box \varphi(x) = 0  \label{wave_equation}
\end{equation}
which is unrelated to Planck's constant $h$ and does not contain a mass or length scale like
the reduced Compton wave length $\lambdabar_c$.
The massless wave equation (\ref{wave_equation}) is used to describe massless scalar particles in flat spacetime,
which, as a remarkable matter of fact, have never been observed experimentally in the
physical particle spectrum. The massless case has to be clearly distinguished from the massive case
due to group theoretical reasons related to the representation theory of the proper orthochronous Poincar\'e group;
there is no classical counterpart to a massless quantum particle.\\

Massless scalar fields play some theoretical r\^ole in cosmological inflation models on curved
spacetime \cite{Inflaton}, however, contrary to the massive case which differs strongly from the massless
case due to the existence of a length scale and the corresponding exponential decay of correlation functions,
the scalar massless case with its special properties has not been discussed in the literature in great detail so far.
One reason might be the fact that massless interacting theories like, e.g., massless scalar electrodynamics, the theory of
the electromagnetic interactions of a mass-zero charged scalar field, has led to
endless paradoxes like infinite cross sections when one tried to interpret it in a consistent manner,
i.e. there is strong evidence against the existence of an S-matrix in such a theory.
However, the Coleman-Weinberg mechanism shows a way out in the sense that
radiative corrections may produce spontaneous symmetry breakdown in theories for which the semiclassical (tree)
approximation does not indicate such breakdown such that the theory of a massless charged scalar
finally becomes the theory of a massive vector meson and a massive scalar meson \cite{Coleman}.
\\

For the sake of notational convenience, above and in the following
Cartesian Minkowski coordinates $x$ are used where the speed of
light {\emph{in vacuo}} $c$ is equal to one such that
$x^\mu=(ct,\vec{x})=(x^0,x^1,x^2,x^3)=(x_0,-x_1,-x_2,-x_3)$ and $\partial_\mu=\partial/ \partial x^\mu$.
Planck's constant plays no r$\hat{\mbox{o}}$le then since there is no first-quantized massless
field theory. The spacetime-dependent phase of a plane wave with four-wave number $k^\mu$,
or four-momentum $p^\mu = \hbar k^\mu$   is then given by
$kx=k_\mu x^\mu = k^0 x^0 -k^1 x^1 -k^2 x^2 -k^3 x^3 = k_0 x^0 - \vec{k} \cdot \vec{x}$.\\

\section{Solutions of the massless wave equation}
For the sake of completeness, we start with some basic considerations concerning
massless scalar fields. In many cases, the generalization of the discussion below
to the massive case is straightforward.\\

The general solution of the wave equation
\begin{equation}
\Box  \varphi(x)=0 \quad \mbox{or} \quad -k^2  \hat{\varphi}(k) =0
\end{equation}
has its support on the light cone in momentum space, since $ \hat{\varphi}(k)=0$ must hold for
$k^2 \neq 0$, and can be written in the form
\begin{equation}
 \varphi(x) = \int \frac{d^4 k}{(2 \pi)^4}  \hat{\varphi}(k) e^{-ikx}
= \int \frac{d^4 k}{(2 \pi)^3} \delta(k^2)  \tilde{\varphi}(k) e^{-ikx}
\end{equation}
With $\omega(\vec{k}) = |\vec{k}| = \sqrt{k_1^2 + k_2^2 + k_3^2}$, the distributional identity
\begin{equation}
\delta(k^2)= \delta((k^0-|\vec{k}|)(k^0+|\vec{k}|)) = \frac{1}{2 \omega(\vec{k})}(\delta(k^0-|\vec{k}|) + \delta(k^0+|\vec{k}|)) \, ,
\end{equation}
holds, hence the frequency decomposition with $k=(\omega(\vec{k}),\vec{k})$
\begin{equation}
 \varphi(x) = 
\int \frac{d^3 k}{(2 \pi)^3 2 \omega(\vec{k})} \bigl[ \hat{\varphi}^- (\vec{k}) e^{-ikx} +  \hat{\varphi}^+ (\vec{k}) e^{ikx} \bigr] \, ,
\end{equation}
follows, with
\begin{equation}
 \hat{\varphi}^+ (\vec{k})= \tilde{\varphi}(-\omega(\vec{k}),-\vec{k}) \, .
\end{equation}
The objects considered above may have a smooth analytic or a singular distributional (operator valued) character,
depending on the theoretical setting under study. For the moment, $ \varphi(x)$ shall be considered as a complex Klein-Gordon wave
function. Since we want to describe single particle states and since we have to work in a Hilbert space setting in quantum
mechanics, the $ \hat{\varphi}^\pm$ shall be elements of the two Hilbert spaces $\hat{\mathds{H}}_0^\pm$ which result from the unique
completion of the space of infinitely differentiable functions of compact support $(C_0^\infty (\mathds{R}^3), \, ( \cdot , \cdot ) )$ equipped
with the Lorentz invariant relativistic scalar product
\begin{equation}
( \hat{\varphi}^\pm ,\hat{\psi}^\pm ) = \int 
\frac{d^3 k}{(2 \pi)^3 2 \omega(\vec{k})}   \hat{\varphi}^{\pm} (\vec{k})^* \hat{\psi}^\pm (\vec{k}) \, , \quad  \label{LIS}
\end{equation}
with the star $^*$ denoting complex conjugation. In configuration space, the wave equation generates a unitary dynamics on both
spaces $\mathds{H}_0^{\pm}$.
Then the expression
\begin{equation}
j^\mu (x) = i  \varphi (x)^*   \! \stackrel {\leftrightarrow}{\partial^\mu}  \!  \varphi(x) =i  \varphi (x)^* \partial^\mu  \varphi(x) + c.c.
\label{cc}
\end{equation}
gives rise to a real four-current Klein-Gordon density
which fulfills the continuity equation $\partial_\mu j^\mu=0$,
and the total charge
\begin{displaymath}
\int \limits_{x^0=const.} d^3 x \, j^0(x) = \int \limits_{x^0=const.} d^3 x \, 
\int \frac{d^3 k'}{(2 \pi)^3 2 \omega(\vec{k}')} \int \frac{d^3 k}{(2 \pi)^3 2 \omega(\vec{k})}
\end{displaymath}
\begin{displaymath}
\times \bigl[  \hat{\varphi}^- (\vec{k}')^* e^{ik'x} +  \hat{\varphi}^+ (\vec{k}')^* e^{-ik'x} \bigr]
\bigl[\omega(\vec{k})   \hat{\varphi}^- (\vec{k}) e^{-ikx} - \omega(\vec{k})  \hat{\varphi}^+ (\vec{k}) e^{ikx} \bigr]
+ c.c.
\end{displaymath}
\begin{displaymath}
=\int \limits_{x^0=const.} \frac{d^3 k}{2(2 \pi)^3 2 \omega(\vec{k})} \bigl[ | \hat{\varphi}^- ( \vec{k}) |^2
- | \hat{\varphi}^+ (\vec{k}) |^2 +  \hat{\varphi}^+ (\vec{k})^*  \hat{\varphi}^- (-\vec{k}) e^{-2i \omega(\vec{k}) x^0}
-  \hat{\varphi}^- (\vec{k})^*  \hat{\varphi}^+ (-\vec{k}) e^{+2i \omega(\vec{k}) x^0}
\bigr]
\end{displaymath}
\begin{displaymath}
+ c.c.
\end{displaymath}
\begin{equation}
= \int \limits_{x^0=const.} \frac{d^3 k}{(2 \pi)^3 2 \omega(\vec{k})} \bigl[ | \hat{\varphi}^- ( \vec{k}) |^2
- | \hat{\varphi}^+ (\vec{k}) |^2 \bigr] \, , \label{current}
\end{equation}
where the distributional Fourier transform of the Dirac $\delta$-distribution
\begin{equation}
\int d^n x \, e^{\pm i ( \vec{k}-\vec{k'}) \cdot \vec{x}} = ( 2 \pi)^n \delta^{(3)} (\vec{k}-\vec{k'})
\label{Foudel}
\end{equation}
for $n=3$ has been used.
The standard interpretation of the indefinite current eq. (\ref{cc}) is that $ \varphi(x)$ describes two
particles with opposite charge: a particle with negative-frequency wave function 
\begin{equation}
\varphi^- (x) =  \int \frac{d^3 k}{(2 \pi)^3 2 \omega(\vec{k})}
\hat{\varphi}^- (\vec{k}) e^{-ikx}  \, ,
\end{equation}
and an anti-particle with positive-frequency wave function
\begin{equation}
\varphi^+ (x) =  \int \frac{d^3 k}{(2 \pi)^3 2 \omega(\vec{k})}  \hat{\varphi}^+ (\vec{k}) e^{+ikx} \, .
\end{equation}
Of course, it is a matter of pure convention to label the frequency of the wave function of a particle as negative.
Both in the negative- and the positive-frequency case, the energy of the (anti-)particle is positive.\\

It is common usage to call the current eq. (\ref{cc}) a local quantity, since it depends
\emph{locally} on quantities of the scalar field like the field strength or its derivatives.
Still, the situation is a bit more involved.

\section{Localization of the massless Klein-Gordon current}
In order to understand the locality properties of the charge current eq. (\ref{cc}) properly,
the structure of the charge density $j^\mu(x)$ has to be understood from both the single particle wave function
aspect as well as from the quantum field theoretical point of view.
A complex Klein-Gordon wave function $\varphi(x)$ can be specified, e.g., by Cauchy data at $t=x^0=0$
\begin{equation}
 \varphi(0,\vec{x})= \varphi_0 (\vec{x}) \, , \quad \dot{\varphi}(0,\vec{x})=\dot{\varphi}_0 (\vec{x})
\end{equation}
for the future $x^0 > 0$.
However, if one considers a wave function describing only one particle alone in the universe, the situation is completely
different, since the time derivative $\partial_0  \varphi(x)= \dot{\varphi}(x)$ depends in this case on the wave function $ \varphi(x)$.
One has, e.g., for a negative-frequency and correspondingly positive-energy one-particle wave function
\begin{equation} \varphi^- (x) = \int \frac{d^3 k}{(2 \pi)^3 2 \omega(\vec{k})}  \hat{\varphi}^- (\vec{k})
e^{i(\vec{k} \cdot \vec{x} -\omega(\vec{k}) \cdot t)} \, ,
\end{equation}
which transforms passively under a change of inertial systems described by
a proper orthochronous Poincar$\acute{\mbox{e}}$-transformation
$x^\mu \rightarrow x'^\mu=\Lambda^\mu_{\, \, \nu}x^\nu  +a^\mu$ or $x'= \Lambda x + a$,
where $a$ is a four-vector and the Lorentz transformation matrix
$\Lambda \in SO^+(1,3)$ fulfills
$\Lambda^\mu_{\, \, \alpha} g_{\mu \nu} \Lambda^\nu_{\, \, \beta}
=g_{\alpha \beta} = \mbox{diag} (1,-1,-1,-1)$ or $\Lambda^T g \Lambda =g$, $\det{A}=1$ and $\Lambda^0_{\, \, 0} \ge 1$,
via
\begin{equation}
{\varphi^-}' (x')= \varphi'^- (x')= \varphi^-(x)= \varphi^- (\Lambda^{-1}(x'-a)) \, .
\end{equation}
Indeed, the time derivative of the wave function can be expressed
by the wave function itself via the \emph{non-local} expression 
\begin{equation}
\dot{\varphi}^-(x) =
\int \frac{d^3 k}{(2 \pi)^3 2 \omega(\vec{k})} (-i \omega(\vec{k}))  \hat{\varphi}^- (\vec{k}) 
e^{i(\vec{k} \cdot \vec{x} -\omega(\vec{k}) \cdot t)} =
-\frac{i}{2} \int \frac{d^3 k}{(2 \pi)^3 } 
 \hat{\varphi}^- (\vec{k}) e^{i(\vec{k} \cdot \vec{x} -\omega(\vec{k}) \cdot t)} \, , \label{nonlocal}
\end{equation}
becoming a convolution in position (or 'configuration') space
\begin{equation}
 \dot{\varphi}^- (x^0, \vec{x}) = \int d^3 x' \,  \sigma(\vec{x}-\vec{x}')  \varphi^- (x^0, \vec{x}')
\label{convo}
\end{equation}
with the integral kernel
\begin{equation}
\sigma(\vec{x}-\vec{x}') = - \frac{i}{(2 \pi)^3} \int d^3 k \, |\vec{k}| e^{i \vec{k} \cdot (\vec{x}-\vec{x}')} \, .
\label{kernel}
\end{equation}
The kernel can be calculated easily by the help of the well-known distributional identity involving the Dirac
delta distribution
\begin{equation}
\Delta_k \frac{1}{| \vec{k}|} = \bigl( \partial_{k^1}^2 +\partial_{k^2}^2 +\partial_{k^3}^2 \bigr) 
\frac{1}{| \vec{k}|} = -4 \pi \delta^{(3)} (\vec{k})
\end{equation}
which can be Fourier transformed to
\begin{equation}
\int d^3 k \,  e^{i \vec{k} \cdot \vec{x}}  \Delta_k | \vec{k} |^{-1} = -4 \pi \int d^3 k   \, \delta^{(3)}  (\vec{k})
 e^{i \vec{k} \cdot \vec{x}} = -4 \pi \, ,
\end{equation}
therefore from shifting the momentum-space Laplace operator by partial integration to the exponential
phase term above one has
\begin{equation}
-|\vec{x}|^2 \int d^3 k \, \frac{e^{i \vec{k} \cdot \vec{x}}}{| \vec{k}|} = -4 \pi 
\end{equation}
and consequently
\begin{equation}
 \int d^3 k \, \frac{e^{i \vec{k} \cdot \vec{x}}}{| \vec{k}|} = \frac{4 \pi}{| \vec{x} |^2} \, .
\end{equation}
In an analogous manner, one may write
\begin{equation}
-| \vec{x} |^2 \sigma(\vec{x}) = - \frac{i}{(2 \pi)^3} \int d^3  k \, e^{i \vec{k} \cdot \vec{x}} \Delta_k |\vec{k}| 
= - \frac{2i}{(2 \pi)^3} \int d^3 k \, \frac{e^{i \vec{k} \cdot \vec{x}}}{| \vec{k} |} \, ,
\end{equation}
where $\Delta_k | \vec{k} | = 2/|\vec{k}|$ has been used,
leading to the result
\begin{equation}
\sigma(\vec{x}) = \frac{i}{\pi^2} |\vec{x}|^{-4} \, .  \label{massless_kernel}
\end{equation}
This highly singular distribution can be expressed in a regularized form
as a derivative in position space of the smoother distribution $| \vec{x} |^{-2}$
\begin{equation}
\frac{1}{2} \Delta \frac{1}{| \vec{x} |^2} = \frac{1}{| \vec{x} |^4} \, .
\end{equation}
Using the distibutional derivatives like
\begin{equation}
\Delta \ln |\vec{x}|^2 = \frac{2}{| \vec{x}|^2}
\end{equation}
the spherically symmetric kernel can be written in dipole or tripole form
\begin{equation}
\sigma (\vec{x}) = \frac{i}{(2 \pi)^2} \Delta^2 \ln | \vec{x}|^2 = \frac{i}{6 (2 \pi)^2} \Delta^3 (x^2 \ln | \vec{x} |^2 ) \, .
\end{equation}
Eq. (\ref{kernel}) can also be evaluated directly (with $k=|\vec{k}|$ and $x=|\vec{x}|$ for notational convenience here)
\begin{displaymath}
\sigma(\vec{x}) = - \frac{i}{(2 \pi)^3} \int \limits_{0}^{\infty} dk \, k^2 \int \limits_{-1}^{+1} 2 \pi \, d  (\cos \vartheta) \,
k e^{ikx \cos \vartheta}
=  - \frac{1}{(2 \pi)^2 x } \int  \limits_{0}^{\infty}  dk \, k^2 (e^{ikx}-e^{-ikx})
\end{displaymath}
\begin{equation}
=  - \frac{i}{2 \pi^2 x } \int  \limits_{0}^{\infty}   dk \, k^2 \sin (kx) \, .
\end{equation}
This divergent integral becomes meaningful as the weak (i.e. distributional) limit of the convergent
integrals
\begin{equation}
\sigma(\vec{x}) =  - \frac{i}{2 \pi^2 x } \lim \limits_{\alpha \searrow 0} \int  \limits_{0}^{\infty} dk \, k^2 \sin (kx) e^{-\alpha k} =
 - \frac{i}{2 \pi^2 x } \lim \limits_{\alpha \searrow 0} \frac{2x ( 3 \alpha^2 -x^2)}{(x^2+\alpha^2)^3} = \frac{i}{\pi^2 x^4} \, ,
\end{equation}
again reproducing eq. (\ref{massless_kernel}).\\

We conclude that the Klein-Gordon density $j^0 (x)$ in the setting of single particle wave mechanics
is only formally a local quantity, since its definition contains a time derivative depending in a non-local
manner on the single particle wave function. It will be shown below in a more general way in what sense 
a single particle cannot be localized.\\

In the case of massive particles, the corresponding integral kernel $\sigma^m(\vec{x})$ of a particle with
mass $m$ involves the Compton wavelength $m^{-1}$ of course, and decays exponentially on this length scale.
The uncertainty in eq. (\ref{convo}) discussed above has nothing to do with the usual quantum mechanical position
uncertainty of a particle due to the spatial extension of the wave function in the non-relativistic case.
In a way, it is a generic uncertainty of the single-particle wave function itself.\\

Some expressions given above can be obtained as special cases from the general distributional weak limit
 given for $g>-1$ by
\begin{equation}
\int \limits_{0}^{\infty} dk \, k^g \sin (kx) = \lim \limits_{\alpha \searrow 0} \int  \limits_{0}^{\infty} dk \, k^g \sin (kx) e^{-\alpha k} =
\cos \bigl( \pi g/2) \Gamma(1+g) x^{-(g+1)} \, .
\end{equation}

\section{Negative charge densities from positively charged particles}
Interpreting $j^0$ generated by a single particle wave $\varphi^- (x)$ function containing one frequency type only
\begin{equation}
\varphi^-(x) = \int \frac{d^3 k}{(2 \pi)^3 2 \omega ( \vec{k} )} \hat{\varphi}^- ( \vec{k} ) e^{-ikx} \, , \quad
k^0= \omega(\vec{k}) = | \vec{k} | \, , 
\end{equation}
has strange consequences, since even in the presence of negative or positive frequencies only $j^0$ can be
positive and negative at different spacetime regions.
This can easily be illustrated by superposing two negative frequency plane waves according to
\begin{equation}
\varphi^-_{sup} (x) = e^{-ik_1 x} + \alpha e^{-ik_2 x} \, , \quad
\omega_1= k_1^0=|\vec{k}_1| \, , \quad \omega_2=k_2^0 = | \vec{k}_2| \, , \quad 
\alpha \in \mathds{R} \, .
\end{equation}
Concentrating on the spatial origin of the coordinate system for the sake of simplicity, one has ($x^0=t$)
\begin{displaymath}
j^0 (t, \vec{0}) = (e^{i \omega_1 t} + \alpha e^{i \omega_2 t}) (\omega_1 e^{-i \omega_1 t} + \alpha
\omega_2 e^{-i \omega_2 t}) + c.c.
\end{displaymath}
\begin{equation}
= 2 [\omega_1 + \alpha^2 \omega_2 + \alpha ( \omega_1 + \omega_2) \cos ( (\omega_1 - \omega_2) t) ] \, .
\end{equation}
E.g., for the numerical values $\omega_1=1$, $\omega_2=4$, and $\alpha=1/2$ one has
$\omega_1 + \alpha^2 \omega_2 = 2 < 5/2 = \alpha (\omega_1 + \omega_2)$, and therefore the
amplitude of the oscillatory cosine-term, which is greater than the temporal mean of $j^0 (t, \vec{0})$,
will cause local oscillations between negative and positive values of $j^0$. Even if the Klein-Gordon
wave function is restricted to a single particle type, the charge density can be indefinite. This is quite a remarkable
result, since the indefiniteness of the Klein-Gordon density is often attributed in the literature to the presence of
both positive and negative frequency solutions of the Klein-Gordon equation. If fact, even when one of the
frequency types is projected away, the Klein-Gordon density remains indefinite.\\

The incorrect claim that for negative-frequency wave functions, the timelike
component $j^0(x)$ of the Klein-Gordon current is positive-definite in Minkowski space, and
that therefore "a consistent theory [of localization in Minkowski space] can be developed
for a free [relativistic spin-zero] particle" \cite{Schweber} turns out to be
totally false since, as eventually rigorously proved by Gerlach et al. \cite{Gerlach1, Gerlach2, Gerlach3}, the
opposite is actually true: for any single-frequency solution of the Klein-Gordon equation
there are points in Minkowski space where $j^0(x) < 0$ and points where $j^0(x) > 0$ at the same
time. Hence, the Klein-Gordon current is never a probability
current. As a matter of fact,
the problem of sharp localization of spin-zero quantum particles in
relation to classical Lorentz frames turns out to be unsolvable, as will be discussed in detail at the end
of this paper. 
This observation remains true for arbitrary spin values, for massive as well as for
quantum particles of zero mass. Claims that no problems
with the conventional notion of particle localizability occur in the case of relativistic
quantum particles of spin-1/2 also turn out to be wrong \cite{Wightman}. 

\section{Non-covariant localization}
Defining a wave function $\hat{\varphi}^-_{NW} (\vec{k})$ in momentum space in the sense of
Newton and Wigner \cite{NewtonWigner}
\begin{equation}
\hat{\varphi}^-_{NW} (\vec{k}) = (2 \omega(\vec{k}))^{-1/2} \hat{\varphi}^- (\vec{k}) \, ,
\end{equation}
the norm squared of a positive energy single particle state becomes, according to the Lorentz invariant scalar
product defined in eq. (\ref{LIS})
\begin{equation}
\int d^3 x \, \hat{\varphi}^-_{NW} (x^0,\vec{x})^*  \hat{\varphi}^-_{NW} (x^0,\vec{x}) =
\int d^3 k \, \hat{\varphi}^-_{NW} (\vec{k})^*  \hat{\varphi}^-_{NW} (\vec{k})
\end{equation}
due to Parseval's theorem, with the Newton-Wigner wave function in position space given by
\begin{equation}
\hat{\varphi}^-_{NW} (x^0,\vec{x}) = \int \frac{d^3 k}{(2 \pi)^3  \sqrt{2 \omega(\vec{k})}} \, 
\hat{\varphi}^- (\vec{k}) e^{-ikx} \, .
\end{equation}
Accordingly, $\hat{\varphi}^-_{NW} (x^0,\vec{x})$ can be written as a convolution
\begin{equation}
\hat{\varphi}^-_{NW} (x^0,\vec{x}) = \int \frac{d^3 k}{(2 \pi)^3 2 \omega({\vec{k})}} \sqrt{2 \omega(\vec{k})}
\hat{\varphi}^- (\vec{k}) e^{-ikx} =
\int d^3 x' \, \sigma_{NW} (\vec{x} - \vec{x}') \varphi^- (x^0, \vec{x}') 
\end{equation}
with
\begin{equation}
\sigma_{NW} (\vec{x} -\vec{x}') = \int \frac{d^3k}{(2 \pi)^3}
\frac{e^{i \vec{k} (\vec{x}-\vec{x}')}}{ \sqrt{2 \omega(\vec{k})}  } \, . \label{NWKernel}
\end{equation}
Using the distributional Fourier transform eq. (\ref{Foudel}), eq. (\ref{NWKernel}) follows from the
short calculation
\begin{displaymath}
\int d^3 x' \sigma_{NW} (\vec{x} - \vec{x}') \hat{\varphi}^- (x^0, \vec{x}')
\end{displaymath}
\begin{displaymath}
=\frac{1}{(2 \pi)^6} \int d^3 x' \int d^3 k \int \frac{d^3 k'}{2 \omega(\vec{k'})} \, \sqrt{2 \omega(\vec{k}')}
e^{i \vec{k}' (\vec{x}-\vec{x}')} \hat{\varphi}^- (\vec{k}) e^{i \vec{k}  \vec{x}' - i \omega(\vec{k}) x^0} 
\end{displaymath}
\begin{equation}
= \frac{1}{(2 \pi)^3} \int d^3 k \int \frac{ d^3 k'}{ \sqrt{2 \omega(\vec{k'})}} \delta^{(3)} (\vec{k}'-\vec{k})
\hat{\varphi}^- (\vec{k}) e^{i \vec{k}'  \vec{x} - i \omega(\vec{k}) x^0} \, .
\end{equation}
The kernel can also be evaluated directly (with $k=|\vec{k}|$ and $x=|\vec{x}|$ for notational convenience)
\begin{displaymath}
\sigma_{NW} (\vec{x}) =  \frac{1}{(2 \pi)^3} \int \limits_{0}^{\infty} dk \, \frac{k^2}{\sqrt{2 k}} \int \limits_{-1}^{+1} 2 \pi \,
d ( \cos \vartheta ) \, e^{ikx \cos \vartheta}
=  - \frac{i}{ \sqrt{2} (2 \pi)^2 x } \int  \limits_{0}^{\infty}  dk \, k^{1/2} (e^{ikx}-e^{-ikx})
\end{displaymath}
\begin{equation}
=   \frac{1}{\sqrt{2} 2 \pi^2 x } \int  \limits_{0}^{\infty}   dk \, k^{1/2} \sin (kx)
=  \frac{1}{8 \sqrt{ \pi^3 }} \frac{1}{x^{5/2}} \, ,
\end{equation}
where the distributional weak limit
\begin{equation}
\int \limits_0^\infty dk \, k^{1/2} \sin (kx) = 
\cos (\pi/ 4) \Gamma (3/2) x^{-3/2} = \frac{\sqrt{2 \pi}}{4}  x^{-3/2} 
\end{equation}
was used.\\


We mention here that the Newton-Wigner wave function $\varphi_{NW} (x)$ is related to the covariant wave function $\varphi(x)$
in the massive case by
\begin{equation}
\varphi_{NW}^m (x^0, \vec{x}) = \frac{1}{(2 \pi)^2 \Gamma (1/4)} \sqrt{\frac{\pi}{2}} \int d^3x'
\biggr( \frac{2}{\lambdabar_c |\vec{x} - \vec{x}'|} \biggr)^{5/4}
K_{5/4} \biggl( \frac{| \vec{x} - \vec{x}'|}{\lambdabar_c} \biggr) \varphi(x^0, \vec{x}' ) \, .
\end{equation}
A full derivation of this result, which is hardly found stated correctly in the literature, is given in the appendix.

\section{Second quantization}
In order to provide a well-defined setting for the forthcoming discussion
on a quantum field theoretical level, we discuss some basic properties and definitions concerning the free, i.e.
non-interacting scalar quantum field describing neutral or charged spin-0 particles of mass $m$ in (3+1) spacetime dimensions.
It is rather common to represent such a field according to
\begin{displaymath}
\varphi(x)=\varphi^{-}(x)+\varphi^{+}(x)
\end{displaymath}
\begin{equation}
=
\frac{1}{(2 \pi)^{3/2}} \int \frac{d^3 k}{\sqrt{2 k^0}}
[a(\vec{k}) e^{-ikx}+a^\dagger(\vec{k}) e^{+ikx}] \quad \mbox{(neutral)},
\end{equation}
\begin{displaymath}
\varphi_c(x)=\varphi_c^{-}(x)+\varphi_c^{+}(x)
\end{displaymath}
\begin{equation}
=
\frac{1}{(2 \pi)^{3/2}} \int \frac{d^3 k}{\sqrt{2 k^0}}
[a(\vec{k}) e^{-ikx}+b^\dagger(\vec{k}) e^{+ikx}] \quad \mbox{(charged)},
\end{equation}
where $k x= k_\mu x^\mu=k^0 x^0 - \vec{k} \cdot \vec{x}$, $k^0 \overset{!}{=}E= \omega(\vec{k})=
\sqrt{\vec{k}^2+m^2}>0$,
$\pm$ denotes the positive and negative frequency parts of the fields and
$\dagger$ a 'hermitian conjugation'. For notational convenience, the same symbol for the quantized field
and its non-quantized version discussed above shall be used.
The non-vanishing \emph{distributional} commutator relations for the destruction and creation field operators
in the above Fourier decomposition are
\begin{equation}
[a(\vec{k}),a^\dagger(\vec{k'})]=[b(\vec{k}),b^\dagger(\vec{k'})]=
\delta^{(3)} (\vec{k}-\vec{k'}) \, , \label{algeb1}
\end{equation}
otherwise
\begin{equation}
[a(\vec{k}),a(\vec{k'})]=[b(\vec{k}),b(\vec{k'})]
=[a^\dagger
(\vec{k}),a^\dagger(\vec{k'})]
=[b^\dagger(\vec{k}),b^\dagger(\vec{k'})]= 0  \label{algeb2}
\end{equation}
and
\begin{equation}
[a(\vec{k}),b(\vec{k'})]=[a(\vec{k}),b^\dagger(\vec{k'})]
=[a^\dagger
(\vec{k}),b(\vec{k'})]=[a^\dagger(\vec{k}),b^\dagger(\vec{k'})]= 0  \label{algeb2a}
\end{equation}
holds.
The destruction (or 'annihilation', or 'absorption')  operators act on the \emph{non-degenerate} vacuum
$|0\rangle$ according to
\begin{equation}
a(\vec{k}) |0 \rangle=b(\vec{k}) |0 \rangle = 0 \quad \mbox{for all}  \, \,  k \in \mathds{R}^3 \, . \label{vacstate}
\end{equation}
It is crucial to require the existence of a state $| 0 \rangle$ which is annihilated by
all the $a(\vec{k})$ and  $b(\vec{k})$, since otherwise there would be many inequivalent irreducible Hilbert space
representations of the algebraic relations given by eqns. (\ref{algeb1}) -(\ref{algeb2a}), and eq. (\ref{vacstate})
selects the one in Fock space where the $a(\vec{k})$ and  $b(\vec{k})$ can be interpreted as destruction and
the $a^\dagger(\vec{k})$ and  $b^\dagger(\vec{k})$ as creation (or 'emission') operators.\\

Single-particle wave functions or states of, e.g., $a$-particles
represented in momentum space $\Psi_1(\vec{k})$, $\Psi_2(\vec{k})$
are
\begin{equation}
| \Psi_1  \rangle = \int d^3 k \, \Psi_1(\vec{k}) a^\dagger(\vec{k})|0 \rangle  \, , \quad 
| \Psi_2 \rangle = \int d^3 k' \, \Psi_2(\vec{k'}) a^\dagger(\vec{k'}) |0 \rangle \, , \label{single_particle}
\end{equation}
their scalar product becomes, from a formal but correct distributional calculation exploiting the commutation relations above,
\begin{displaymath}
\langle \Psi_1 | \Psi_2 \rangle =
\int d^3 k  d^3 k' \, {\Psi_1}(\vec{k})^* \Psi_2(\vec{k'})\langle 0 | a(\vec{k}) a^\dagger(\vec{k'}) | 0 \rangle
\end{displaymath}
\begin{equation}
=\int d^3 k  d^3 k' \, {\Psi_1}(\vec{k})^* \Psi_2(\vec{k'}) \langle 0 | [\delta^{(3)}(\vec{k}-\vec{k'}) +
a^\dagger(\vec{k'}) a(\vec{k}) ] | 0 \rangle
= \int d^3 k  \, {\Psi_1}(\vec{k})^* \Psi_2(\vec{k})  \, .
\end{equation}
This scalar product can be written in a manifestly covariant form by using differently normalized
creation and destruction operators fulfilling
\begin{equation}
[\tilde{a}(\vec{k}),\tilde{a}^\dagger(\vec{k'})]=[\tilde{b}(\vec{k}),\tilde{b}^\dagger(\vec{k'})]=
(2 \pi)^3 (2 k^0) \delta^{(3)} (\vec{k}-\vec{k'}) \, .
\end{equation}
Then one represents the single $a$-particle states by
\begin{equation}
| \Psi_1  \rangle = \frac{1}{( 2 \pi)^3} \int \frac{d^3 k}{2 \omega(\vec{k})}  \, \tilde{\Psi}_1(\vec{k})
\tilde{a}^\dagger(\vec{k})|0 \rangle  \, , \quad
| \Psi_2 \rangle = \frac{1}{( 2 \pi)^3} \int \frac{d^3 k'}{2 \omega(\vec{k'})}  \, \tilde{\Psi}_2(\vec{k'})
\tilde{a}^\dagger(\vec{k'}) |0 \rangle \, , 
\end{equation}
and the scalar product becomes
\begin{equation}
\langle \Psi_1 | \Psi_2 \rangle
= \int \frac{d^3 k}{( 2 \pi)^3 2 \omega(\vec{k})}  \, {\tilde{\Psi}_1}(\vec{k})^* \tilde{\Psi}_2(\vec{k}) 
= \int \frac{d^4 k}{(2 \pi)^3} \delta(k^2-m^2) \Theta(k^0) \hat{\Psi}_1 (k)^* \hat{\Psi}_2 (k) 
\end{equation}
with $\hat{\Psi}_{1,2} (\sqrt{\vec{k}^2+m^2}, \vec{k})= \tilde{\Psi}_{1,2} (\vec{k})$.
$\Theta$ denotes the Heaviside step distribution.\\

In configuration space, the two strategies just described are directly related to the description of a particle by
a Klein-Gordon wave function or the corresponding Newton-Wigner wave function.
The appealing property of the normalization according to eq. (\ref{algeb1}) is the fact that after an inverse
three-dimensional Fourier transform, the configuration space operator
\begin{equation}
a(\vec{x}) = \frac{1}{(2 \pi)^{3/2}} \int d^3k \, a(\vec{k}) e^{i \vec{k} \vec{x}}
\end{equation}
implies the seemingly local commutation relation
\begin{equation}
[a(\vec{x}) , a^\dagger (\vec{x}')] = \delta^{(3)} ( \vec{x} - \vec{x}') \, .
\end{equation}
However, the Newton-Wigner type wave function of a particle state $| \vec{x} \rangle = a^\dagger (\vec{x}) | 0 \rangle$
'created at a point' $\vec{x}$
\begin{displaymath}
\langle 0 | \varphi(x') | \vec{x} \rangle =
\frac{1}{(2 \pi)^3} \langle 0 | \int \frac{d^3 k'}{\sqrt{2 k'^0}} a(\vec{k'}) e^{-i k' x'}
\int d^3 k \, a^\dagger ( \vec{k}) e^{-i \vec{k} \vec{x}} | 0 \rangle
\end{displaymath}
\begin{equation}
= \frac{1}{(2 \pi)^3} \int \frac{d^3 k}{\sqrt{2 k^0}} e^{-i k^0 x'^0 + i \vec{k} (\vec{x}' - \vec{x})} 
\end{equation}
has no point-like support.

\section{Causal properties of commutation and correlation distributions}
From the above algebraic relations represented by free fields on a Fock space $\mathcal{F}$
one constructs the scalar Feynman propagator as  distributional time-ordered vacuum expectation value
\begin{equation}
\Delta_F(x-y)=-i \langle 0 | T (\varphi_c (x) \varphi_c^\dagger(y)) | 0 \rangle \, ,
\end{equation}
where translational invariance implies
\begin{equation}
\Delta_F(x)=-i \langle 0 | T (\varphi_c (x) \varphi_c^\dagger(0)) | 0 \rangle
\end{equation}
or
\begin{equation}
\Delta_F(x)=-i \langle 0 | T (\varphi(x) \varphi(0)) | 0 \rangle,
\end{equation}
for neutral fields.
The wave equation holds in a distributional sense
\begin{equation}
(\Box+m^2) \Delta_F(x)=(\partial_\mu \partial^\mu+m^2) \Delta_F(x)=- \delta^{(4)} (x)
\end{equation}
and one also defines the positive- and negative-frequency Pauli-Jordan $C$-number distributions
or, up to an imaginary factor, 'Wightman two-point functions'
\begin{equation}
\Delta^{\pm}(x)=-i[\varphi^{\mp}(x),\varphi^{\pm}(0)] =-i[\varphi_c^{\mp}(x),\varphi_c^{\dagger \, \pm}(0)]  \, ,
\end{equation}
\begin{equation}
\Delta(x)=\Delta^{+}(x)+\Delta^{-}(x)
= -i [\varphi(x),\varphi(0)] = -i [\varphi_c(x), \varphi_c^\dagger(0)]\, ,
\end{equation}
i.e.
\begin{displaymath}
\Delta^{+}(x)=-i \langle 0 | \varphi^- (x) \varphi^+ (0) | 0 \rangle \, ,
\end{displaymath}
\begin{equation}
\Delta^{-}(x)=+i \langle 0 | \varphi^- (0) \varphi^+ (x) | 0 \rangle \, . \label{delta_plus}
\end{equation}
The retarded propagator is given by $\Delta^{ret}(x)=\Theta(x^0) \Delta(x)$, a product of distributions
which is well-defined due to the harmless scaling behaviour of $ \Delta(x)$ at the origin $x=0$.\\

Some important properties of the objects and their Fourier transforms introduced so far are enlisted
in the following:
$\Delta(x)$ vanishes for space-like arguments $x$ with $x^2 < 0$, as required by causality. One has
\begin{displaymath}
\hat{\Delta}^{\pm}(k)=\frac{1}{(2 \pi)^2} \int d^4 x \, \Delta^{\pm} (x) e^{ikx}
\end{displaymath}
\begin{equation}
=\mp \frac{i}{2 \pi} \Theta(\pm k^0)
\delta(k^2-m^2) \, , \label{pj_expl}
\end{equation}
\begin{equation}
\Delta^+(x) = - \Delta^-(-x) \, \ ,
\end{equation}
\begin{equation}
\Delta(x)=\Delta^+(x)-\Delta^+(-x) \, ,
\end{equation}
\begin{equation}
\Delta(-x)=-\Delta(x) \, .
\end{equation}
\begin{equation}
\Delta_F(x)=\Theta(x^0) \Delta^+(x) - \Theta(-x^0) \Delta^-(x) \, .
\end{equation}
\begin{equation}
(\Box+m^2) \Delta^{\pm}(x)=0 \, ,  \quad (k^2-m^2) \hat{\Delta}^\pm(k)=0 \, .
\end{equation}
\begin{equation}
\Delta^{ret}=\Delta_F+\Delta^{-} \, ,
\end{equation}
\begin{equation}
\displaystyle \Delta^{ret}(x)=\int \frac{d^4 k}{(2 \pi)^4} \frac{e^{-ikx}}{k^2-m^2+i k^0 0} \, ,
\end{equation}
\begin{equation}
(\Box+m^2) \Delta^{ret}(x)=- \delta^{(4)} (x) \, .
\end{equation}
For $m=0$ the scalar Feynman propagator in configuration space is
\begin{displaymath}
\Delta_F^0(x)
=\int \frac{d^4 k}{(2 \pi)^4} \frac{e^{-ikx}}{k^2+i0}
\end{displaymath}
\begin{equation}
=\frac{i}{4 \pi^2} \frac{1}{x^2-i0} =
\frac{i}{4 \pi^2} P \frac{1}{x^2}-
\frac{1}{4 \pi} \delta (x^2) \,  ,
\end{equation}
where $P$ denotes the principal value and  $\delta$ the one-dimensional  Dirac distribution, and
the massless Pauli-Jordan distributions in configuration space are given by
\begin{equation}
\displaystyle \Delta^0(x)
=-\frac{1}{2 \pi} \mbox{sgn} (x^0) \delta(x^2) \,  ,
\end{equation}
\begin{equation}
\displaystyle  \Delta^\pm_0(x) = \pm \frac{i}{4 \pi^2} \frac{1}{(x_0 \mp i0)^2 -\vec{x}^2} \, ,
\end{equation}
and since $\Delta^{ret}(x)=\Theta(x^0) \Delta(x)$ one has
\begin{equation}
\Delta^{ret}_0(x)
=-\frac{1}{2 \pi} \Theta (x^0) \delta(x^2) \, .
\end{equation}

A notational issue concerning the principal value in the case of  $\Delta_0^+$ is clarified by
\begin{displaymath}
\frac{1}{(x^0-i0)^2-\vec{x}^2}=\frac{1}{((x^0-i0)-|\vec{x}|)((x^0-i0)+|\vec{x}|)}
\end{displaymath}
\begin{equation}
=\frac{1}{2|\vec{x}|} \frac{1}{x^0-|\vec{x}|-i0}- \frac{1}{2|\vec{x}|} \frac{1}{x^0+|\vec{x}|-i0}
=P \frac{1}{x^2}+i \pi \mbox{sgn}(x^0) \delta(x^2)
\end{equation}
or
\begin{equation}
\frac{1}{(x^0-i0)^2-\vec{x}^2}=
\frac{1}{x^2-2i0 x^0-0^2}=\frac{1}{x^2-i0 \mbox{sgn}(x^0)}
= P \frac{1}{x^2}+i \pi \mbox{sgn}(x^0) \delta(x^2) \, .
\end{equation}

\section{Locality and causality in quantum field theory}
As a matter of fact, causality is not completely understood from a philosophical and physical point of view.
Technically, in quantum field theory causality is usually expressed in the form of (anti-)commu\-ta\-tion relations
for bosonic (fermionic) operator valued distributions which are intimately
connected with the support properties of the corresponding objects. In quantum field theory,
quantum field operators are local, but states are non-local.
This has far reaching consequences for renormalization techniques used in perturbative
quantum field theory \cite{Aste}.\\

A single particle theory is problematic due to its non-local aspects, but in quantum field theory
particles together with their anti-particles conspire in such a way that causal propagation
of specific physically relevant quantities is ensured and can be calculated by the help of, e.g.,
integral kernels like the retarded propagator $\Delta^{ret}$, which has its {\emph{causal}}
distributional support
\begin{equation}
\mbox{supp} \, \Delta^{ret} (x) \subseteq \overline{V}^+ = \{ x \mid x^2 \ge 0 \, , \, x^0 \ge 0 \}
\end{equation}
in the closed forward (future-directed) light-cone $\overline{V}^+$,
i.e. $\Delta^{ret} (\varphi) = 0$
holds for all test functions in the Schwartz space $\varphi \! \in
\! \mathcal{S}(\mathds{R}^4)$ with support $\mbox{supp} ( \varphi ) \subset
\mathds{R}^4 - \overline{V}^+$.\\

A \emph{particular} solution of the inhomogeneous Klein-Gordon equation
\begin{equation}
(\Box  + m^2) \varphi(x) = j(x) \, , \label{inhomo_sol}
\end{equation}
where the source term $j$ may act locally and have compact support in spacetime, is given by
\begin{equation}
\varphi^{part}(x) = - \int d^4 x' \, \Delta^{ret} (x-x') j(x') \, ,
\end{equation}
since 
\begin{equation}
(\Box + m^2)  \int d^4 x' \, \Delta^{ret} (x-x') j(x') = - \int d^4 x' \,
\delta^{(4)} (x-x') j(x') = -j(x) \, .
\end{equation}
Causal behaviour of $\varphi^{part}$ is ensured by eq. (\ref{inhomo_sol}) in the sense that
$\mbox{supp} \, \varphi^{part}$ will lie in the causal future of $\mbox{supp} \, j$, but it will necessarily
contain negative and positive frequency parts.\\

The Pauli-Jordan distribution $\Delta$ also has {\emph{causal support}}, it vanishes outside the closed
forward light-cone and backward light-cone $\overline{V}^-$ such that
\begin{equation}
\mbox{supp} \, \Delta(x) \subseteq \overline{V}= \overline{V}^- \cup \overline{V}^+  \, , \quad
\overline{V}^-=\{x \, | \, x^2 \ge 0, \,  x^0 < 0 \}
\end{equation}
in the sense of distributions. Furthermore, it
solves the wave equation $\Box \Delta(x) = 0$ with the Cauchy data
$\Delta(0,\vec{x})=0$ and $(\partial_0 \Delta)(0, \vec{x})= -\delta^{(3)}(\vec{x})$.
This specific feature of $\Delta$ that it can be restricted as a distribution to a space-like hyperplane
allows the calculation of a Klein-Gordon wave function from Cauchy data given, e.g., at $x^0=ct=0$
\begin{equation}
\varphi_0(\vec{x})=\varphi(0,\vec{x}) \, , \quad \dot{\varphi}_0(\vec{x}) = \partial_0 \varphi(x) |_{x^0=0} \, .
\end{equation}
From the homogeneous solution with $(\Box +m^2) \varphi(x)=0$
\begin{equation}
\varphi(x)= - \int \limits_{x'^0=0} d^3 x' \, \Delta (x-x') \,  \overset{\leftrightarrow}{\partial'_0} \, \varphi(x')
\label{general_sol}
\end{equation}
follows indeed (note that $\partial'_0 \Delta(x-x') = - \partial_0 \Delta(x-x')$)
\begin{displaymath}
\varphi(0,\vec{x}) = - \int d^3 x' \, \bigl[ (\partial_0 \Delta) (0, \vec{x}-\vec{x}') \varphi_0(\vec{x}')
+ \Delta(0,\vec{x}-\vec{x}') \dot{\varphi}(0,\vec{x}') \bigr]
\end{displaymath}
\begin{equation}
= \int d^3 x' \delta^{(3)} (\vec{x}- \vec{x}') \varphi_0(\vec{x}') = \varphi_0 (\vec{x}) 
\end{equation}
and
\begin{displaymath}
(\partial_0 \varphi) (0,\vec{x}) = - \int d^3 x' \, \bigl[ (\partial_0^2 \Delta) (0, \vec{x}-\vec{x}') \varphi_0(\vec{x}')
+ (\partial_0 \Delta) (0,\vec{x}-\vec{x}') \dot{\varphi}(0,\vec{x}') \bigr]
\end{displaymath}
\begin{equation}
= \int d^3 x' \delta^{(3)} (\vec{x}- \vec{x}') \dot{\varphi}_0(\vec{x}') = \dot{\varphi}_0 (\vec{x}) \, ,
\end{equation}
since all time derivatives of even order of $\Delta(x^0,\vec{x})$ restricted to $x^0=0$ vanish.
Solution eq. (\ref{general_sol}) contains both frequency types
when the Cauchy data $\varphi_0$ and $\dot{\varphi}_0$ have compact support
on the hyperplane defined by $x^0=0$.
In analogy to eq. (\ref{general_sol}) one may construct 'causal' Klein-Gordon waves
\begin{equation}
\varphi^{ret}(x)= - \int \limits_{x'^0=0} d^3 x' \, \Delta^{ret} (x-x') \,  \overset{\leftrightarrow}{\partial'_0} \, \varphi(x')
\end{equation}
which, however, still contain particle and anti-particle frequencies.
Singling out one frequency type in eq. (\ref{general_sol}) according to
\begin{equation}
\varphi^\mp(x)=- \int \limits_{x'^0=0} d^3 x' \, \Delta^\pm (x-x') \,  \overset{\leftrightarrow}{\partial'_0} \, \varphi(x')
\end{equation}
leads to solutions which do not respect Einstein causality in the sense that the waves $\varphi^\pm$ do not propagate in the
causal future of $\mbox{supp} \, \varphi_0 \cup \mbox{supp} \, \dot{\varphi}_0$ when these supports are compact sets
in the hyperplane $x^0=0$. This is due to the acausal support properties of the $\Delta^{\pm}$-distributions, which
lead to a further astonishing observation.\\

From
\begin{equation}
\langle 0 | T ( \varphi(x) \varphi(0) ) | 0 \rangle = \Theta(x^0) [\varphi(x),\varphi(0)] +
\langle 0 | \varphi(0) \varphi(x) | 0 \rangle
\end{equation}
follows
\begin{equation}
{\Delta}_F(x)={\Delta}^{ret}(x)-{\Delta}^-(x)
\end{equation}
or
\begin{equation}
{\Delta}^-(x)={\Delta}^{ret}(x)-{\Delta}_F(x) \, .
\end{equation}
It follows that the negative-frequency Pauli-Jordan distribution
\begin{equation}
\Delta^{-}_0 (x)=-\frac{1}{2\pi}\Theta(x^0) \delta(x^2) - \frac{i}{4 \pi^2} \frac{1}{x^2-i0} 
\end{equation}
does not vanish for space-like arguments
\begin{equation}
\Delta^{-}_0 (x)= - \frac{i}{4 \pi^2} \frac{1}{x^2} \, ,  \quad  x^2<0 \, ,
\end{equation}
i.e., $\Delta^- (x)$ has no causal support. Since by definition
\begin{equation}
\langle 0 | \varphi(0) \varphi(x) | 0 \rangle = \langle 0 | \varphi^- (0) \varphi^+ (x) | 0 \rangle 
= - i \Delta^{-}_0 (x)
\end{equation}
or
\begin{equation}
\langle 0 | \varphi(x_1) \varphi(x_2) | 0 \rangle = - \frac{1}{4 \pi^2} \frac{1}{(x_1-x_2)^2}
\end{equation}
for $(x_1-x_2)^2 <0$, a stunning observation can be made when one considers two wave functions
$\psi_1(x)$ and $\psi_2(x)$ with two disjoint compact, causally separated supports
\begin{equation}
(x_1-x_2)^2 < 0 \, \, \mbox{for all} \, \, x_1 \in \mbox{supp} \, \psi_1 \, , \, \,
x_2 \in \mbox{supp} \, \psi_2 \, .
\end{equation}
Calculating the overlap of the single particle states
\begin{equation}
| \Psi_1 \rangle = \int d^4 x_1 \, \psi_1 (x_1) \varphi(x_1) | 0 \rangle = \int  d^4 x_1 \, \psi_1 (x_1) \varphi^+ (x_1) | 0 \rangle\, ,
\end{equation}
\begin{equation}
| \Psi_2 \rangle = \int d^4  x_2 \, \psi_2 (x_2) \varphi(x_2) | 0 \rangle = \int d^4 x_2 \,  \psi_2 (x_2) \varphi^+ (x_2) | 0 \rangle  \, ,
\end{equation}
the result
\begin{equation}
\langle \Psi_1 | \Psi_2 \rangle = - \frac{1}{(2 \pi)^2} \int d^4 x_1 \int d^4 x_2 \,  
\frac{\psi_1 (x_1)^* \psi_2(x_2)}{ (x_1-x_2)^2}  \neq 0 \, 
\end{equation}
turns out to be non-vanishing in general although the particles are created in causally disconnected spacetime regions.
This observation is intimately related with the famous Reeh-Schlieder theorem \cite{Reeh}.\\

To conclude this paper, a rigorous discussion of the conflict between the single-particle locality and causality shall
be presented following the lines given in \cite{Hegerfeldt}.
To this end, one considers a single-particle state represented by a single-frequency Klein-Gordon wave function
$\varphi(x)$ which is supposed to be localized in a compact three-dimensional space region $V(0)$ at the initial time $x^0=ct=0$.
Given the initial data in configuration or momentum space
\begin{equation}
\varphi(x^0=0, \vec{x}) = \int \frac{d^3 k}{(2 \pi)^3 2 k^0} \, \hat{\varphi} (\vec{k}) e^{i \vec{k} \vec{x}} \, ,
\end{equation}
the wave function evolves according to
\begin{equation}
\varphi(x) = \int \frac{d^3 k}{(2 \pi)^3 2 k^0} \,
\hat{\varphi}(\vec{k}) e^{i \vec{k} \vec{x} - i k^0 x^0} = U(t) \varphi(0, \vec{x}) \, .
\end{equation}
with $k^0=\sqrt{\vec{k}^2 + m^2}$ and particle mass $m \ge 0$, defining implicitly the propagator $U(t)$.
As observed above, the Klein-Gordon density $j^0$ cannot serve as a probability density.
In order to give a precise meaning to the term 'localization', the assumption
is made that some operator $P_{V(0)}$ exists such that the expectation value
$\langle \psi | P_{V(0)} | \psi \rangle$ represents the probability to find the particle in the region $V(0)$ for any
particle state represented by an arbitrary Klein-Gordon wave function $\psi$.
For the following, it indeed suffices to consider one single region $V(0)$ only.
As a quantum mechanical probability operator, $P_{V(0)}$ should be hermitian and fulfill
\begin{equation}
0 \le \langle \psi | P_{V(0)} | \psi \rangle \le 1 \, .
\end{equation}
If $\varphi(x^0=0,\vec{x})$ is localized in $V(0)$, it is an eigenstate
of $P_{V(0)}$ with
\begin{equation}
\langle \varphi(0) | P_{V(0)} | \varphi(0) \rangle =1 \, , \quad
P_{V(0)} | \varphi(0) \rangle = | \varphi(0) \rangle \, .
\end{equation}
\begin{figure}
\begin{center}
\includegraphics[width=8.5cm]{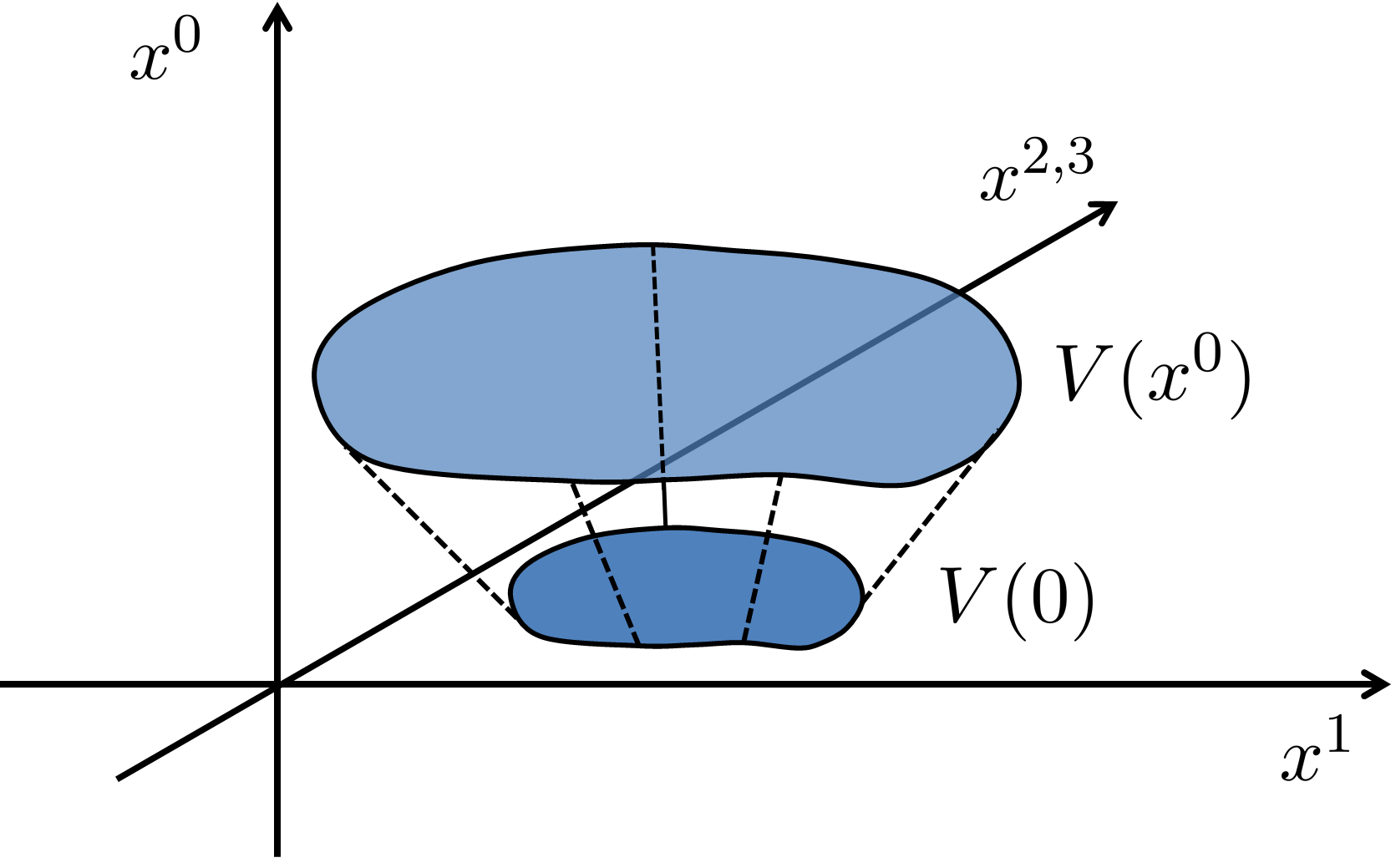}
\caption{3-dimensional spacelike localization volumes at two different times $x^0=ct$.}
\label{loca_vol}
\end{center}
\end{figure}

In order to invoke causality in the sense that no signal can propagate faster than the speed of light,
one observes that the probability to find a particle which was localized at $x^0=0$ in
$V(0)$ should vanish at a later time $x^0 >0$ outside the bounded region ($\vec{x} \sim (0, \vec{x})$)
\begin{equation}
V(x^0) = \{ \vec{x} \mid \mbox{dist} (\vec{x}, V(0)) \le x^0=ct \}
\end{equation}
where
\begin{equation}
\mbox{dist} (\vec{x}, V(0)) = \mbox{inf} \, \{ \mbox{dist} (\vec{x}, \vec{y}) \mid \vec{y} \in V(0) \} \, ,
\end{equation}
with the Euclidean distance
\begin{equation}
\mbox{dist} (\vec{x}, \vec{y})=\sqrt{(x_1-y_1)^2+(x_2-y_2)^2+(x_3-y_3)^2}= || \vec{x} - \vec{y} || \, ,
\end{equation}
as depicted in Figure \ref{loca_vol}.
On the other side, considering a translated wave function
\begin{equation}
U(\vec{a}) \varphi(x^0, \vec{x}) = \varphi(x^0, \vec{x}+\vec{a}) 
\end{equation}
for some $\vec{a}$ with $|| \vec{a}|| \ge r(ct)$ for a sufficiently large constant $r(x^0)$ at the time $x^0>0$,
one must have
\begin{equation}
\langle U(\vec{a}) U(t) \varphi(0) | P(V) | U(\vec{a}) U(t) \varphi(0) \rangle = 0 \, ,
\label{ueberlapp}
\end{equation}
since a sufficiently dislocated wave function at $t=0$ cannot reach the region $V(x^0)$ at $t=x^0/c$.
It is well-known that $P_{V(0)}$ possesses a self-adjoint square root, so eq. (\ref{ueberlapp}) implies
\begin{equation}
P_{V(0)}^{1/2} U(\vec{a}) U(t) |\varphi(0) \rangle = 0 \, \, \rightarrow \, \, 
P_{V(0)} U(\vec{a}) U(t) |\varphi(0) \rangle = 0 \, .
\end{equation}
The scalar product
\begin{equation}
\langle \varphi(0) | P_{V(0)} U(\vec{a}) U(t) | \varphi(0) \rangle = 0
\end{equation}
vanishes therefore and consequently for $|| \vec{a} || \ge r(x^0)$
\begin{equation}
\int \frac{d^3 k}{\sqrt{\vec{k}^2+ m^2}} | \hat{\varphi} (\vec{k}) |^2 e^{-i \sqrt{\vec{k}^2 + m^2} x^0}
e^{+i \vec{k} \vec{a}} = 0 \label{paley}
\end{equation}
vanishes also.
According to Schwartz's Paley-Wiener theorem, the Fourier transform of a distribution with
compact support is an \emph{entire} function
\cite{Constantinescu}, and since the expression in eq. (\ref{paley}) is quite well-behaved, one may conclude
that
\begin{equation}
s(\vec{k},x^0)=\frac{1}{\sqrt{\vec{k}^2+ m^2}} | \hat{\varphi} (\vec{k}) |^2 e^{-i \sqrt{\vec{k}^2 + m^2} x^0}
\end{equation}
is a holomorphic function in $\mathds{C}^3$ for, e.g., $x^0=0$, since the Fourier transform of $s(\vec{k},0)$
has a compact support, but this is true also for arbitrary times $x^0>0$. However, due to the square root in the exponent
containing the time, $s(\vec{k},x^0)$ cannot be entire for two distinct times.\\

Of course,
subtle questions which already bothered Pauli \cite{Pauli}
persist in interacting quantum field theory when one realizes that measuring local quantities like quantum field strength
expectation values requires {\emph{test particles}} which are not strictly localizable.

\section{Appendix: Massive Newton-Wigner distribution}
The Newton-Wigner integral kernel is given in the massive case
by
\begin{align}
\sigma_{NW}^m  (\vec{x}) & = \int \dfrac{\mathrm{d}^{3} k}{(2 \pi)^{3}} \dfrac{\mathrm{e}^{\mathrm{i} \vec{k} \cdot
\vec{x}}}{\sqrt{2 \omega(\vec{k})}}
\\
& = \dfrac{1}{\sqrt{2} (2 \pi)^{3}} \int \mathrm{d}^{3} k \, \dfrac{\mathrm{e}^{\mathrm{i} \vec{k}
\cdot \vec{x}}}{(k^{2} + m^{2})^{1/4}}
\\
& = \dfrac{2 \pi}{\sqrt{2} (2 \pi)^{3}} \int_{0}^{\infty} \mathrm{d} k \, \dfrac{k^{2}}{(k^{2} + m^{2})^{1/4}}
\int_{-1}^{+1} \mathrm{d}(\cos \theta) \mathrm{e}^{\mathrm{i} k x \cos \theta}
\end{align}
with $k = \vert \vec{k} \vert$, $x = \vert \vec{x} \vert$ and $\omega(\vec{k}) = \sqrt{ \vec{k}^{2} + m^{2} }$).
From 
\begin{align}
\int_{-1}^{+1} \mathrm{d}(\cos \theta) \mathrm{e}^{\mathrm{i} k x \cos \theta} & =
\dfrac{1}{\mathrm{i} k x} \left( \mathrm{e}^{i k x} - \mathrm{e}^{- i k x} \right)
= \dfrac{2 \sin(k x)}{k x}
\end{align}
one obtains
\begin{align}
\sigma_{NW}^m (\vec{x}) & = \dfrac{4 \pi}{\sqrt{2} (2 \pi)^{3}} \dfrac{1}{x} \int_{0}^{\infty}
\mathrm{d} k \, \dfrac{k \sin(k x)}{(k^{2} + m^{2})^{1/4}}
\\
& = - \dfrac{4 \pi}{\sqrt{2} (2 \pi)^{3}} \dfrac{1}{x} \dfrac{\mathrm{d}}{\mathrm{d} x}
\int_{0}^{\infty} \mathrm{d} k \, \dfrac{\cos(k x)}{(k^{2} + m^{2})^{1/4}} \, .
\end{align}

Besset's integral (8.432 eq. (5) in \cite{Gradshteyn}),
\begin{equation} \label{Basset}
K_{\nu}(x z) = \dfrac{\Gamma(\nu + \frac{1}{2}) (2 z)^{\nu}}{\sqrt{\pi} x^{\nu}} \int_{0}^{\infty}
\mathrm{d}t \, \dfrac{\cos(x t)}{(t^{2} + z^{2})^{\nu + \frac{1}{2}}} \, ,
\quad \mathrm{Re} \,  \nu
> - \dfrac{1}{2}, x> 0 \, , \quad \vert \mathrm{arg} \, z \vert < \dfrac{\pi}{2}
\end{equation}
 with $\nu = -1/4$ and $z = m$ leads to

\begin{equation}
\int_{0}^{\infty} \mathrm{d} k \, \dfrac{\cos(k x)}{(k^{2} + m^{2})^{1/4}} =
\dfrac{\sqrt{\pi} x^{-1/4}}{\Gamma(1/4) (2 m)^{-1/4}} K_{-1/4}(m x) \, ,
\end{equation}
so one can write
\begin{equation}
\sigma_{NW}^m (\vec{x}) = - \dfrac{4 \pi \sqrt{\pi} (2 m)^{1/4}}{\sqrt{2} (2 \pi)^{3}
\Gamma(1/4)} \dfrac{1}{x} \dfrac{\mathrm{d}}{\mathrm{d} x}\left( x^{-1/4} K_{-1/4}(m x) \right).
\end{equation}

$ $
\\
From the derivative relation 8.486 eq. (14) in \cite{Gradshteyn} follows
\begin{equation}
\dfrac{1}{z} \dfrac{\mathrm{d}}{\mathrm{d} z} \left( z^{\nu} K_{\nu}(z) \right) = - z^{\nu -1} K_{\nu - 1}(z) \, ,
\label{Bessel_diff}
\end{equation}
hence
\begin{equation}
\dfrac{1}{x} \dfrac{\mathrm{d}}{\mathrm{d} x} \left( x^{\nu} K_{\nu}(m x) \right) = - m x^{\nu -1} K_{\nu - 1}(m x),
\end{equation}
and therefore, with $\nu = -1/4$ and $\nu - 1 = -5/4$
\begin{equation}
\sigma_{NW}^m (\vec{x}) = \dfrac{4 \pi \sqrt{\pi} m (2 m)^{1/4}}{\sqrt{2} (2 \pi)^{3} \Gamma(1/4)}
\dfrac{1}{x^{5/4}} K_{-5/4}(m x) \, .
\end{equation}

$ $
\\
Finally, from the
connection formula for order index $\nu$ (8.486 eq. (16) in \cite{Gradshteyn})
\begin{equation} \label{Bessel_index}
K_{- \nu}(z) = K_{\nu}(z)
\end{equation}
and the corresponding relation $K_{-5/4}(m x) = K_{5/4}(m x)$ the desired result follows
\begin{equation}
\sigma_{NW}^m (\vec{x}) = \dfrac{1}{(2 \pi)^{2} \Gamma(1/4)} \sqrt{\dfrac{\pi}{2}}
\left( \dfrac{2 m}{x} \right)^{5/4} K_{5/4}(m x) \, ,
\end{equation}
which can be expressed in terms of the reduced Compton wave length $\lambdabar_{c} = 1/m$ ($\hbar = c = 1$)
\begin{equation}
\sigma_{NW}^m (\vec{x}) = \dfrac{1}{(2 \pi)^{2} \Gamma(1/4)} \sqrt{\dfrac{\pi}{2}}
\left( \dfrac{2}{\lambdabar_{c} x} \right)^{5/4} K_{5/4}(m x) \, .
\end{equation}

The small argument limit (eq. (9.6.9) in \cite{Abramovitz})
\begin{equation} \label{Bessel_lim}
K_{\nu}(z) \overset{z \to 0}{\sim} \dfrac{1}{2} \Gamma(\nu) \left( \dfrac{1}{2} z \right)^{-\nu},
\quad \mathrm{Re} \, \nu > 0
\end{equation}
leads to
\begin{equation}
K_{5/4}(m x) \overset{m \to 0}{\sim} \dfrac{1}{2} \Gamma(5/4) \left( \dfrac{1}{2} m x \right)^{-5/4} \, ,
\end{equation}
and together with the functional equation 
\begin{equation}
\Gamma(x+1) = x \Gamma(x)
\end{equation}
the massless kernel is recovered from the massive case
\begin{equation}
\sigma_{NW} (\vec{x}) = \dfrac{1}{8 \sqrt{\pi^{3}}}\dfrac{1}{x^{5/2}} \, .
\end{equation}

\end{document}